# Lithium niobate microring with ultra-high *Q* factor above 10$^8$

**Renhong Gao,**[1,6] **Ni Yao,**[2] **Jianglin Guan,**[3,4] **Li Deng,**[3,4] **Jintian Lin,**[1,6,10] **Min Wang,**[3,4] **Lingling Qiao,**[1] **Wei Fang,**[5] and **Ya Cheng,**[1, 3,4, 6,7,9 ,11]

[1] *State Key Laboratory of High Field Laser Physics and CAS Center for Excellence in Ultra-Intense Laser Science, Shanghai Institute of Optics and Fine Mechanics (SIOM), Chinese Academy of Sciences (CAS), Shanghai 201800, China.*
[2] *Research Center for Intelligent Sensing, Zhejiang Lab, Hangzhou 311100, China.*
[3] *XXL—The Extreme Optoelectromechanics Laboratory, School of Physics and Electronic Science, East China Normal University, Shanghai 200241, China.*
[4] *State Key Laboratory of Precision Spectroscopy, East China Normal University, Shanghai 200062, China.*
[5] *State Key Laboratory of Modern Optical Instrumentation, College of Optical Science and Engineering, Zhejiang University, Hangzhou 310027, China.*
[6] *Center of Materials Science and Optoelectronics Engineering, University of Chinese Academy of Sciences, Beijing 100049, China.*
[7] *Collaborative Innovation Center of Extreme Optics, Shanxi University, Taiyuan 030006, China.*
[8] *Collaborative Innovation Center of Light Manipulations and Applications, Shandong Normal University, Jinan 250358, China.*
[9] *Shanghai Research Center for Quantum Sciences, Shanghai 201315, China.*
[10] *jintianlin@siom.ac.cn*
[11] *ya.cheng@siom.ac.cn*

**Abstract:** We demonstrate ultra-high quality (*Q*) factor microring resonators close to the intrinsic material absorption limit on lithium niobate on insulator. The microrings are fabricated on pristine LN thin-film wafer thinned from LN bulk via chemo-mechanical etching without ion slicing and ion etching. A record-high *Q* factor up to 10$^8$ at the wavelength of 1550 nm is achieved because of the ultra-smooth interface of the microrings and the absence of ion-induced lattice damage, indicating an ultra-low waveguide propagation loss of ∼0.28 dB/m. The ultra-high *Q* microrings will pave the way for integrated quantum light source, frequency comb generation, and nonlinear optical processes.

## 1. Introduction

Lithium-niobate-on-insulator wafer (LNOI) is considered as an important candidate platform for photonic integrated circuits (PICs), owning to its outstanding material properties that include a broad transparency window (350 nm to 5 μm), a linear electro-optic effect, and a large second-order nonlinearity susceptibility ($\chi^{(2)}$=30 pm/V) [1-5]. A wide of high-performance device applications, such as waveguide/resonator optical frequency convertors, high-speed Mach-Zehnder modulators, and multiplexers have been demonstrated due to the rapid developments in ion-slicing technique and LNOI nanofabrication technology [6-18]. Among the various devices, optical waveguides with ultra-low propagation loss and high refractive index contrast are building elements for the realization of large-scale PICs [19-23]. And the propagation losses in optical waveguides are susceptible to sidewall roughness induced by the fabrication imperfection. To reduce the sidewall roughness of the waveguide, diamond-blade dicing, precision cutting, focused ion beam milling, argon ion milling, and chemo-mechanical etching have been successively used to etch the LNOI into waveguides with smoother sidewalls, resulting in propagation losses ranging from 120 dB/m, below 10 dB/m, to 2.7 dB/m [19-30].

However, the minimum propagation loss of the waveguide determined by the intrinsic material absorption is only 0.1 dB/m, which shows great potential for wafer-scale integration.

In addition to the sidewall-roughness scattering, some other factors, particularly the ion-induced lattice damage caused by ion-slicing and ion milling should also be taken into consideration to further reduce the propagation loss [1,4,31]. To avoid ion-induced lattice damage, chemo-mechanical polishing/etching is proposed to produce the LNOI microdisk. The $Q$ factor as high as $1.23 \times 10^8$ at 1550 nm wavelength, which is close to the intrinsic material absorption limit, has been reported [32]. The microring resonators, compared with microdisk resonators, have great advantages due to their superior characteristics such as flexibility in dispersion engineering, capability of electrical integration, and potential for large-scale integration. However, lithium niobite (LN), as a difficult-to-cut material, is difficult to be structured into a ridge waveguide with low loss.

Here we challenge the status quo and show that the LN ridge waveguides can be fabricated with a propagation loss as low as 0.28 dB/m through suppressing the fabrication imperfection during thin film producing and etching, which is one order lower than reported previously. The loaded $Q$ factor of the fabricated microring is measured as $0.8 \times 10^8$, corresponding to an intrinsic $Q$ factor of $10^8$. Remarkably, the microring loss is close to the intrinsic material absorption limit of LN crystal.

## 2. Fabrication methods

The manufacturing process to fabricate the LN microrings by femtosecond laser direct writing and the chemo-mechanical polishing (CMP) is schematically illustrated in Fig.1. First, the X-cut LN crystal was bonded to the silica buffer layer deposited on another LN bulk crystal, and the thicknesses of the silica layer and bulk crystals were 2 μm and 500 μm, respectively. Second, the top bulk crystal was thinned into 4 μm-thickness thin film via CMP considering the trade-off between the surface evenness and thickness as shown in Ref. 32] . Thus, combined with the techniques of step 1 and step 2, an LNOI wafer was formed. Third, to protect the underneath LN in the CMP process, a 600 nm-thickness chromium (Cr) layer was coated on the LNOI wafer by magnetron sputtering. Forth, the Cr layer was ablated into a microring-pattern hard mask by femtosecond laser direct writing. Fifth, the sample was endured with CMP to etch the exposed LN thin film, leading to the pattern transferring from the Cr layer to LN thin film [33]. Sixth, chemical etching was performed to remove the Cr mask. Finally, a secondary CMP was carried out for thinning the microrings. The main advantage of our fabrication method relies on the optimized process of CMP for thinning LN crystal, alleviating the crystal damage and surface damage.

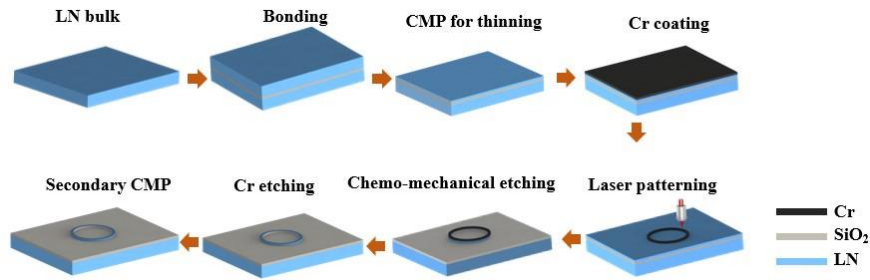

Fig. 1. Illustration of the fabrication flow of the mirorings.

## 3. Characteristics of LN microrings

The optical microscope image and the magnified scanning-electron-microscope (SEM) image of the microrings are shown in Fig. 2(a) and Fig. 2(b), respectively, indicating the LN microring with a diameter of 200 μm is fabricated. To accurately measure the wedge angle and the height of the microrings, a small slit is cut with focused ion beam, as shown in Fig. 2(c), indicating that the wedge angle and the height of the microrings are 9° and 720 nm, respectively. Interestingly, the small wedge angle will drive the modes far from the edge of the microrings, benefiting higher $Q$ factors [30,34]. As a cost of small wedge angle, the bottom with of the microring is relatively large, which is 20 μm. The waveguide width can be further decreased by replacing the thin film wafer with a thinner one.

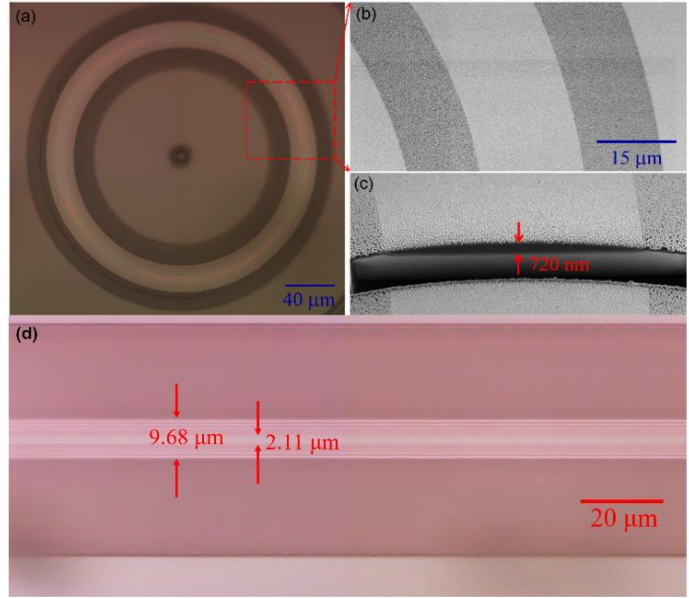

Fig. 2. (a) Optical microscope image of the fabricated mirroring. (b) Magnified scanning-electron-microscope (SEM) image of the fabricated microring. (c) The SEM image shows a small slit is cut with focused ion beam, demonstrating the height of the fabricated microrings. (d) The optical microscope image of the ridge waveguide on other LNOI chip for coupling of the microrings.

Ridge waveguide with top width, bottom width, and height of 2.11 μm, 9.68 μm, and 700 nm, respectively, is fabricated on a second LNOI chip (produced by ion slicing) to couple with the microrings as shown in Fig. 2(d). The experimental setup is schematically illustrated in Fig. 3(a). The ridge waveguide was adjusted to be parallel with the top surface of the microrings by an xyz-θx-θy-rotatability stage, and come into contact with the microring to gain optimum coupling as shown in Fig. 3(b). Lensed fibers were used to couple the light signal into and out of the ridge waveguide by end-fire coupling with a coupling efficiency of 10%/facet. A narrow-linewidth tunable laser with a linewidth of 500 Hz (model: TLB-6728, New focus Inc.) was used as light source with effective in-coupled power of 5 μW (with a variable optical attenuator (VOA)) in the microring to avoid the thermal and nonlinear optical effects [24,29]. And an inline fiber polarization controller (PC) was used to adjust the polarization of the light. The output optical signal was coupled out of the microring by the ridge waveguide and lensed fiber, and sent into a photodetector (PD, model: 1811, New focus Inc.). The transmission spectrum was real-time analyzed by an oscilloscope (model: Tektronix MDO04) when scanning the wavelength of the optical signal. Whispering gallery modes were excited when the optical signal was resonant with the microring, resulting in a spectrum of sharp dips in the transmission spectrum.

Figure 3(c) shows the transmission spectra of the microring at the wavelength ranging from 1564 nm to 1569 nm, indicating an FSR of 2.12 nm. The $Q$ factors of the modes labeled with cross and star in Fig. 3(c) were characterized by scanning the tunable laser around the central wavelengths of either mode, as shown in Fig. 3(d) and Fig. 3(e). The loaded $Q$ factor of the mode at ~1567.7 nm is measured as $2.1\times10^7$, giving an intrinsic $Q$ factor of $2.2\times10^7$ if the mode transmission is considered. The mode at ~1568.6 nm shows a mode splitting with two loaded

$Q$ factors of $5.9\times10^7$ and $8.0\times10^7$ near the condition of critical coupling. Considering the transmission rates of modes, the intrinsic $Q$ factors ($Q_i$) are $0.85\times10^8$ and $1.08\times10^8$, respectively, agreeing well with the result of LNOI microdisks [32]. Remarkably, the highest intrinsic $Q$ factor close to the material intrinsic absorption limit of lithium niobite is about one order of magnitude higher than the best-reported values in LNOI microrings [19,33]. And the corresponding waveguide propagation loss $L_P$ is 0.28 dB/m, which is calculated by:

$$L_p = 10\cdot\alpha\cdot\log e = 10\cdot(2\pi n_{eff}/(Q_i\lambda))\cdot\log e \qquad (1)$$

where $\alpha$ is the material absorption coefficient, $n_{eff}$ is the effective refractive index of the mode, $Q_i$ and $\lambda$ are the intrinsic $Q$ factor and resonant wavelength of the mode, respectively.

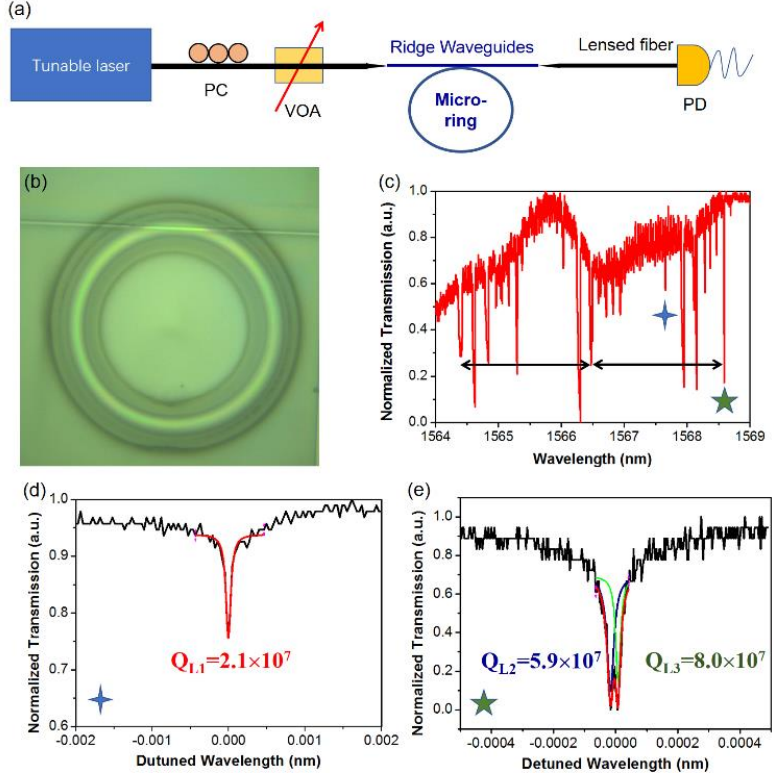

Fig. 3. (a) Experimental setup for mode characterization. (b) Optical microscope image of the waveguide coupled microring. (c) Transmission spectrum around the pump wavelength. (d) and (e) $Q$ factors of the modes labeled in Fig. 3(c).

## 4. Conclusion

In conclusion, we demonstrate an ultra-high $Q$ microring resonator in LNOI without ion-induced lattice damage by CMP. The intrinsic $Q$ factor above $10^8$ was experimentally demonstrated at 1550 nm wavelength band, while the waveguide propagation loss is only 0.28 dB/m. Such a low-loss LNOI ridge waveguide shows great potentials for loss-less electro-optic modulators [31], high-speed information processing, and large-scale photonic integration on LNOI wafer.